\documentclass[a4paper,12pt]{article}
\usepackage{graphicx}

\usepackage[english]{babel}
\usepackage{amsmath}
\usepackage{amssymb}

\begin{document}

\title{Formation of a mesa shaped
   phonon pulse  in  superfluid $^4$He}

\author{I.N. Adamenko, K.E. Nemchenko, and V.A. Slipko \\
V.N. Karazin Kharkov National University, 61077, Ukraine\\
}
\maketitle E-mail:
valery.slipko@gmail.com

\begin{abstract}
We present a theory for the formation of a mesa shaped
phonon pulse  in  superfluid $^4$He.
Starting from  the hydrodynamic equations of superfluid helium, we
obtain the system of
equations which describe the  evolution
of strongly anisotropic phonon systems. Such systems can be created
experimentally.
The solution of the equations are simple waves, which correspond to
second sound in the moving phonon pulse. Using
these exact solutions, we describe the expansion of phonon pulses in
superfluid helium at zero
temperature.
This theory gives an explanation for
the mesa shape  observed in the measured phonon angular distributions.
Almost all dependencies of the mesa shape on the system parameters can
be qualitatively understood.
  \end{abstract}

\section{Introduction}
It is at present, impossible to
create in superfluid helium, a constant relative velocity, ${\bf w}$,
between the normal and superfluid which is close to
the Landau critical velocity.
However  phonon pulses can be created,  which are pulses of normal fluid moving
through the stationary superfluid with a high relative velocity,
without superfluidity breaking down
  \cite{Wyatt1,Wyatt2,Wyatt3,Wyatt4,Wyatt5}. These pulses are
$t_p=10^{-5}-10^{-7}$s, long.

In experiments \cite{Wyatt1}-\cite{Wyatt5}
a heater is immersed in superfluid helium at $\sim$50
mK. A current pulse, applied to the heater, creates a moving phonon
system in the stationary superfluid.
The initial transverse and longitudinal dimensions of such a phonon
system is determined
by the heater size and the duration of the current pulse
respectively. Then the phonon
system starts to expand as it moves in the superfluid helium.
There are two very different regimes of such evolution. At pressure above
19 bar phonons propagate almost without scattering. So there is ballistic
propagation. In this case, the  phonon energy has an angular
distribution with a characteristic
cosine dependence.
At lower pressures, the phonons strongly interact  by three phonon
processes. These are prohibited by the
phonon energy-momentum conservation laws at pressure above 19 bar.
Therefore at low pressures
there can be a hydrodynamic regime, when phonon system expands
in "phonon vacuum" in a way similar to  the ordinary expansion of a
gas into a vacuum.

In a  recent paper \cite{Wyatt5} detailed measurements of the angular
distributions of the energy in the phonon pulse were reported at zero pressure.
The measurements were made at different distances from the heater,
with various heater sizes and heater powers.
The results showed a notable feature; the angular dependence of the
energy flux has a flat top, and concave steep sides.
In \cite{Wyatt5}  this kind of distribution was called a mesa shape.
Such a phonon energy distributions cannot occur in
the ballistic regime. In this case we would have cosine-like convex angular distribution, as it follows both from experiments at high enough pressure and from theoretical considerations (Lambert's law).  At zero pressure we have, despite the assumptions made in the theoretical model in [5], a hydrodynamic regime due to three phonon processes in the wings of the angular distribution, where the phonon density is sufficiently high, as well as in the centre of the pulse.
The dependencies of the mesa width and  height
on various  parameters, obtained in \cite{Wyatt5}, are nontrivial.
So the hydrodynamic propagation of phonon
systems in superfluid helium is an important problem.

To  choose a method of theoretically analysing the phonon pulse expansion
in superfluid helium, we should take into account that in
experiments \cite{Wyatt2,Wyatt3,Wyatt4,Wyatt5}
the phonon pulse duration is much greater than the time to attain a
local equilibrium.
Therefore there is a dynamic equilibrium which arises
   mainly due to scattering by three phonon processes, which have a
characteristic time $t_{3pp}\sim 10^{-8}$s
\cite{3ppTime}. This applies to the experimental conditions
\cite{Wyatt2,Wyatt3,Wyatt4,Wyatt5}.
Fast relaxation   has been observed directly in the
experiments with colliding phonon pulses \cite{hotline1}. There is
further  evidence from the fact that a phonon pulse
propagates in superfluid helium as a whole, with a velocity which is
experimentally indistinguishable from the Landau critical velocity for phonons.
For phonons with a linear energy-momentum relation,
$\varepsilon(p)=cp$, the Landau critical velocity is equal to $c$
\cite{Andreev1}.

Thus the strong three-phonon scattering, within a pulse comprised of low
energy phonons in liquid helium at 0 bar,
leads to a quasi equilibrium.
The equilibrium state of the phonons  can be defined in terms of a
temperature and a drift velocity,
which means  we can use a
hydrodynamic approach to describe  the  dynamics of the phonon pulse.

Another feature of such phonon systems is that the energy density in
the phonon pulse
is much larger than the ambient phonon energy density in superfluid helium.
Therefore the evolution of  a strongly anisotropic phonon  system is
essentially nonlinear.
Nonlinear waves in superfluid helium, when $w$ is small, have been
studied for many
years, but the analysis of nonlinear
waves at arbitrary $w$ has not
been done until now.
The first such analysis was made in Ref.  \cite{EvolConusTheory}, where the
phonon pulse evolution were studied by using the Bose-cone model for
the phonon distribution function.
There,  the nonlinear equations of gas-dynamic type
  for the parameters of the Bose-cone model were obtained.
These equations were solved for one dimensional transverse and
longitudinal evolution of a phonon pulse. In \cite{JLTP2005}
and \cite{JPCM2005} the more reasonable Bose-Einstein phonon
distribution function was used instead
of the Bose-cone distribution. But  the approximate theories
developed in these papers only allowed us
to study phonon evolution which was near to the initial state
(see \cite{JPCM2005}).
The question of the mesa shaped angular distribution did not even
occur in  \cite{JLTP2005} and \cite{JPCM2005}, because
this phenomenon was only fully investigated recently \cite{Wyatt5}.

In the present paper we   rigorously solve the problem of  phonon
pulse expansion into the "superfluid vacuum" of helium.
This gives a  physical explanation of the mesa shape formation and
the nontrivial
dependencies of the mesa width and height on various parameters.

In Sec. II we obtain the nonlinear equations, which describes the  evolution
of  phonon systems created by thermal pulses  in superfluid helium. In Sec.
III we find explicitly the family of exact solutions of these
equations. They are the
simple waves, which correspond to the second sound modes in phonon systems.
In Sec. IV these exact solutions  are used
to describe the first stage of the expansion of a phonon  layer in superfluid
helium, when only the incident waves exist. Here we find the
expansion velocity of a phonon pulse
into the "phonon vacuum", i.e. into superfluid helium with zero temperature.
The second stage of the expansion of the phonon  layer, when  reflected
waves are formed, are studied in Sec. V. Here we develop an
approximate method to describe the reflected wave,
  which allows us to find the average energy density and the width of
the reflected wave.
In Sec. VI by using the theory presented in this paper
we  give  qualitative explanations of some experimental data from Ref.
(\cite{Wyatt5}).
Conclusions are drawn in Sec. VII.

\section{Equations for the evolution of a phonon system in  the
hydrodynamic approximation}

To describe the evolution of a phonon system in superfluid helium, we
start from the
well-known two-fluid hydrodynamic equations in the non-dissipative
approximation \cite{KhalatnikovBook,Putterman}:
\begin{equation}
\frac{\partial \rho}{\partial t}+ div (\rho_n {\bf v}_n +
\rho_s{\bf v}_s)=0; \label{MassEq}
\end{equation}
\begin{equation}
\frac{\partial S}{\partial t}+ div (S {\bf v}_n )=0; \label{EntropyEq}
\end{equation}
\begin{equation}
\frac{\partial {\bf v}_s}{\partial t}+\nabla \mu + ({\bf v}_s
\nabla){\bf v}_s=0; \label{vsEq}
\end{equation}
\begin{equation}
\frac{\partial A_i}{\partial t}+ ({\bf v}_n\nabla)A_i=
-\frac{\partial T}{\partial x_i}-A_k\frac{\partial
v_{nk}}{\partial x_i}. \label{AEq}
   \end{equation}
Here $\rho$ is the density of helium; $\rho_n$ and $\rho_s$ are
densities of the normal and superfluid components respectively; $\bf
v_n$ and $\bf v_s$ are velocities of normal and superfluid
components respectively; $S$ is the entropy of unit of volume; $\mu$
is the chemical potential of unit of mass of helium; ${\bf A}
=\rho_n {\bf w} / S$; ${\bf w} = {\bf v}_n-{\bf v}_s$ is the relative
velocity of the normal and superfluid components; $T$ is temperature.
Summation
over twice repeated indices $k$ is implied in Eq. (\ref{AEq}).
Eqs. (\ref{MassEq}) and (\ref{EntropyEq}) express
the conservation of  mass and entropy respectively, Eq. (\ref{vsEq})
expresses the acceleration of the superfluid and Eq. (\ref{AEq})  is a
combination of momentum conservation and other conservation laws.

In the experiments of Refs. \cite{Wyatt1}-\cite{Wyatt5}, phonon
systems were created by
a heater immersed in superfluid helium. The temperature of the liquid
helium was $\sim$ 50
mK. At this temperature, the ambient thermal excitations can be neglected.
The power of the applied heat pulses was such that the normal density $\rho_n$
was very small, $\rho_n\ll \rho$.
In this case the superfluid velocity and variations of pressure can
be neglected
  when we study phonon propagation \cite{SecSoundPRB}.
Taking into account the approximate relation  ${\bf v}_n = {\bf
w}+{\bf v}_s \approx {\bf w}$ in Eqs. (\ref{EntropyEq}) and
(\ref{AEq}), we obtain the system of equations which describes the
evolution of a phonon system:
\begin{equation}
\frac{\partial S}{\partial t}+ div (S{\bf w} )=0; \label{SEq}
\end{equation}

\begin{equation}
\left(\frac{\partial}{\partial t}+ ({\bf
w}\nabla)\right)\left(\frac{\rho_n}{S}{\bf w}\right)=
-\nabla T-\frac{\rho_n}{2S}\nabla w^{2}.
\label{wEq}
\end{equation}

Relations (\ref{SEq}) and (\ref{wEq}) are the system of four equations for four
variables, for example, for temperature $T$ and the components of
relative velocity ${\bf w}$.
Eqs. (\ref{SEq}) and (\ref{wEq}) must be completed by the
equations of state of superfluid helium $S=S(T,w^2)$,
$\rho_n=\rho_n(T,w^2)$, in which   we have neglected the pressure
dependence in accordance with the assumptions made
above.

The system of Eqs. (\ref{SEq})-(\ref{wEq}) are linearized for small
deviations of the variables.
They define the propagation of
second sound
waves and the transverse wave,   at arbitrary values of the
equilibrium relative velocity
$w$, when $\rho_n\ll\rho$. The dispersion relation for this case is
studied in \cite{SecSoundPRB}.

For a phonon system with a linear energy-momentum relation $\varepsilon=cp$,
where $c^2=(\partial P)/(\partial \rho)$ and $c$ is the first sound velocity of
helium, we have
\cite{KhalatnikovBook}:
\begin{equation}
\frac{\rho_n}{S}=\frac{T}{c^2-w^2},~~S=\frac{2\pi^2}{45}\frac{k_B^4T^3}{\hbar^3c^3(1-w^2/c^2)^2}.
\label{StateEqs}
\end{equation}

The system of equations (\ref{SEq})-(\ref{wEq}), together with the
relations (\ref{StateEqs}), describe
the evolution of phonon systems propagating in superfluid helium.

The main feature of phonon systems, created in experiments
\cite{Wyatt1}-\cite{Wyatt5},
is the high value of the  relative velocity $w$, which has a value
close to the first sound velocity $c$.
At the same time the temperature of such
phonon systems is very small, so the normal density satisfies the
strong inequality
$\rho_n\ll\rho $.
As shown in \cite{PhononStability}, using the general conditions of
stability  for superfluid
helium \cite{Andreev1},  such phonon systems
are thermodynamically stable,
even for values of the relative velocity which approach
the first sound velocity $c$, if the temperature is low enough.

\begin{figure}[t]
\begin{center}
\includegraphics[height=2.5in]{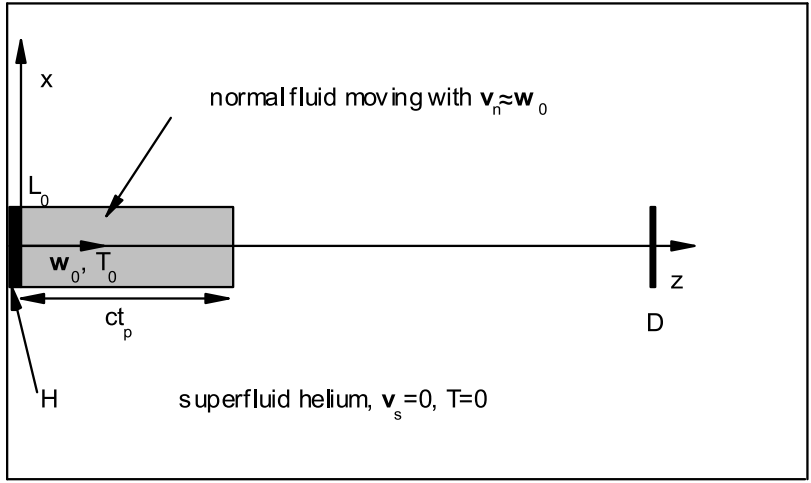}
\end{center}
\caption{A phonon system (shaded region) with characteristic
temperature $T_0$ and
the relative velocity ${\bf w}_0$  in superfluid helium with
  ${\bf v}_s=0$ and temperature $T=0$.
The heater H  has a width $2L_0$.
The characteristic length $ct_p$ of the system depends on the the
pulse duration $t_p$.
The coordinate frame is defined with the $z$-axis parallel to the
normal to the heater H.
D is the detector.}
\label{fig1}
\end{figure}

In a recent paper \cite{Wyatt5} detailed measurements of angular
distributions of energy in phonon systems were reported for
different distances from the heater, and different heater sizes and powers.
Fig.\ref{fig1} illustrates the main idea of the experiments \cite{Wyatt5}.
The phonon system (shaded region)
is created by  a current pulse in the heater H, which is immersed in
superfluid helium.

The created phonon system moves in the direction normal to the heater
(axis $z$) with a velocity close to the first sound velocity $c$. The
pulse expands
transversely. This expansion along with the initial phonon energy distribution
near the heater, determines  the angular distribution of energy
on the detector. By changing the heater size, one can vary the initial
transverse size $L_0$ of the phonon system, and by the changing
heater pulse duration
$t_p$ one can vary the characteristic length $ct_p$ of the system.

We consider the simple case when the dependence on $z$  can be neglected
in the initial value problem for Eqs. (\ref{SEq}), (\ref{wEq}). That
is valid for sufficiently
long  phonon pulses. To study the behaviour of the transverse
expansion of phonon pulses,
we consider  the 2 dimensional case, where all values depend only on
one spatial cartesian coordinate, $x$. The
relative velocity ${\bf w}$ lies in the plane $xz$, i.e. ${\bf w}=(w_x,0,w_z)$.

Substituting Eqs. (\ref{StateEqs}) into Eqs. (\ref{SEq}), (\ref{wEq}),
and introducing the dimensionless variables
\begin{equation}
\Theta=\ln\frac{T}{1-u^{2}},~~u_x=w_x/c,~~u^2=(w_x^2+w_z^2)/c^{2},
\label{uDef}
\end{equation}
we can rewrite the equations in the matrix form
\begin{equation}
\frac{\partial}{c\partial t}
  \left (
    \begin{array}{cc}
    \Theta\\
    u_x\\
    u^2\\
    \end{array}
    \right )
+M(u_x,u^2)\frac{\partial}{\partial x}
\left (
    \begin{array}{cc}
    \Theta\\
    u_x\\
    u^2\\
    \end{array}
    \right )
=0,
\label{MainEqs}
\end{equation}
where the matrix $M(u_x,u^2)$ is equal to
\begin{equation}
\frac{1}{3-u^2}
\left (
    \begin{array}{ccc}
(5-3u^2)u_x & 1-u^2 & -u_x \\
(1-u^2)(3-u^2-2u_x^2) & 2u_x & -0.5(3-u^2-2u_x^2) \\
6(1-u^2)^2 u_x & -2u^2(1-u^2) & 2u^2u_x \\
  \end{array}
    \right ).
\label{MDef}
\end{equation}
We have introduced the
dimensionless relative velocity ${\bf u}={\bf w}/c$ in Eqs.
(\ref{uDef})-(\ref{MDef}).

We see that Eqs.
(\ref{MainEqs})-(\ref{MDef}) is a complicated nonlinear system of
first order equations in partial derivatives, with
coefficients that  depend explicitly on the variable $u^2$ and linearly and
quadratically on the variable $u_x$. We show below that
Eqs. (\ref{MainEqs})-(\ref{MDef})  can be solved exactly for
arbitrary values of the relative velocity $w$.

\section{Second sound simple waves in phonon systems}
In this section  we will obtain the class of exact solutions for  the
system of Eqs.
(\ref{MainEqs})-(\ref{MDef}). In Sec. 4 we apply the solutions, along
with the initial conditions, to the problem
of phonon pulse expansion in superfluid helium with zero temperature.

Let us suppose that all three variables    of the system
of Eqs. (\ref{MainEqs})-(\ref{MDef}) depend on one unknown function,
for example, $\nu=\nu(x,t)$. In other words we seek solution in the form of
$\Theta=\Theta(\nu)$, $u_x=u_x(\nu)$, and $u^2=u^2(\nu)$. Such solutions
are called  simple waves (see, for example, \cite{LL}). Substituting
these relations into
equations  (\ref{MainEqs}), and dividing the equations (\ref{MainEqs})
by $\partial \nu /\partial x$  we obtain the following matrix equation
\begin{equation}
M\frac{d}{d\nu}
\left (
    \begin{array}{cc}
    \Theta\\
    u_x\\
    u^2\\
    \end{array}
    \right )
=-\frac{\frac{\partial\nu}{c\partial t}}
{\frac{\partial\nu}{\partial x}}\frac{d}{d\nu}
\left (
    \begin{array}{cc}
    \Theta\\
    u_x\\
    u^2\\
    \end{array}
    \right ).
\label{EigenEq}
\end{equation}
We see
from Eq. (\ref{EigenEq}) that the vector with components
$(d\Theta/d\nu,du_{x}/d\nu,du^{2}/d\nu)$
is an eigenvector of the matrix $M$, which corresponds to the eigenvalue
$-\frac{\partial\nu}{c\partial t}/\frac{\partial\nu}{\partial x}$.
  Thus, if we know some eigenvector $r=(r_1,r_2,r_3)$ of the matrix
$M$ (\ref{MDef}),
which corresponds to the eigenvalue $\tilde V$, i.e. if
$\sum_{k=1}^3 M_{ik}r_k=\tilde V r_i$ for any index $i\in (1,2,3)$, then
from Eq. (\ref{EigenEq})
we obtain the  running wave equation for the function
$\nu(x,t)$
\begin{equation}
\frac{\partial \nu(x,t)}{c\partial t}+\tilde V(\nu)\frac{\partial
\nu(x,t)}{\partial x}=0,
\label{nuEq}
\end{equation}
which has velocity $c\tilde V$. The system of ordinary differential equations
for this solution is
\begin{equation}
\frac{d\Theta}{d\nu}=\lambda r_{1},~~\frac{du_x}{d\nu}=\lambda
r_{2},~~\frac{du^2}{d\nu}=\lambda r_{3} ,
\label{ODEqs0}
\end{equation}
where $\lambda$ is some multiplier. Below we will show that matrix
$M$ (\ref{MDef})
has three pairwise different eigenvalues. Therefore the eigenvectors,
which correspond
to the same
  eigenvalue, are  collinear, and  $\lambda$ is a coefficient of proportionality
between  a given eigenvector $r=(r_{1},r_{2},r_{3})$ and the
eigenvector with components
$(d\Theta/d\nu,du_{x}/d\nu,du^{2}/d\nu)$.

Eliminating $\lambda$ from the system (\ref{ODEqs0}), we get  the
system of ordinary differential equations
\begin{equation}
\frac{d\Theta}{r_1}=\frac{du_x}{r_2}=\frac{du^2}{r_3},
\label{ODEqs}
\end{equation}
where $r=(r_1,r_2,r_3)$ is a eigenvector  of matrix (\ref{MDef})
of system (\ref{MainEqs}), which corresponds to the eigenvalue $\tilde V$.
The equations (\ref{ODEqs})
determine the functional dependence between the
variables $\Theta$, $u_x$, and $u^2$ in the corresponding simple wave solution
of system (\ref{MainEqs}), (\ref{MDef}).
It follows from
Eqs. (\ref{nuEq}) and (\ref{ODEqs0}), that  any of the variables
$\Theta$, $u_x$, and $u^2$
satisfy the same running wave equation (\ref{nuEq}) as the function $\nu(x,t)$.

The  three eigenvalues of matrix (\ref{MDef}) are calculated to be
\begin{equation}
\tilde V_{1}=\frac{2u_x - \sqrt{(1-u^{2})(3-u^{2}-2u_{x}^2)}}{3-u^2},
\label{V1}
\end{equation}
\begin{equation}
\tilde V_{2}=\frac{2u_x+ \sqrt{(1-u^{2})(3-u^{2}-2u_{x}^2)}}{3-u^2},
\label{V2}
\end{equation}
\begin{equation}
\tilde V_{3}=u_x.
\label{V3}
\end{equation}
It should be noted that the eigenvalues (\ref{V1})-(\ref{V3})
of matrix (\ref{MDef}) of system of Eqs. (\ref{MainEqs}) are real, if $u<1$,
  and $\tilde V_1<\tilde V_3<\tilde V_2$. Thus the system of Eqs.
(\ref{MainEqs}) are
hyperbolic. In this connection it is interesting to note, that the
condition $w<c$,
which guaranties  hyperbolicity of the system  of Eqs.
(\ref{MainEqs}),  (\ref{MDef})
coincides here with the condition of thermodynamic stability for a
phonon system
\cite{PhononStability}.

The expressions (\ref{V1})-(\ref{V3}) give the   simple wave velocities
$V_{1,2,3}=c \tilde V_{1,2,3}$
in a phonon system. Two of these velocities coincide, as they should,
with those of second
sound propagation  in a phonon system \cite{SecSoundPRB} and the third with the
velocity of transverse waves \cite{TransversePRB}. In this case, the
values $u_x$ and $u^2$ in Eqs. (\ref{V1})-(\ref{V3}) are the mean
constant equilibrium
values, from which there are
small perturbations of temperature and
relative velocity, which  propagate with velocities (\ref{V1})-(\ref{V3}).
Velocities (\ref{V1}) and (\ref{V2})  correspond to  the second sound
propagation velocities, which were
found in \cite{SecSoundPRB} for arbitrary values of relative velocity $w$.
  Particularly, at $w=0$, we obtain  $V_{1,2}=\mp c/\sqrt{3}$
which is the well-known phonon second sound velocity, when
$\rho_n\ll\rho$.  Specific
features of second sound in anisotropic phonon systems with $w\neq0$,
i.e. when the system is characterised by a certain direction of the
relative velocity
${\bf w}$, were studied in  \cite{SecSoundPRB}.

The simple wave velocity  (\ref{V3}) corresponds to the one for the
so-called transverse
wave with the dispersion law $\omega={\bf k v_n}\approx {\bf k w}=kw_x$.
The properties of the transverse wave and possibility of realising it in
phonon pulses, were discussed in \cite{TransversePRB}.

Let us consider the simple second sound wave
with velocity (\ref{V1}).
The corresponding eigenvector
of matrix (\ref{MDef}) is equal to
\begin{equation}
r=\begin{pmatrix}r_1 \\
r_2 \\
r_3\\
\end{pmatrix}
=
\begin{pmatrix}2u_x-(1-u^2)R \\
(1-u^2)(u_x+R)R \\
2(1-u^2)(3u_x+u^2 R)\\
\end{pmatrix},
\label{r}
\end{equation}
where we denote
\begin{equation}
R=\sqrt{\frac{3-u^2-2u_x^2}{1-u^2}}.
\label{R}
\end{equation}
Substituting the second and third components of the eigenvector (\ref{r})
into the system of equations (\ref{ODEqs}), we obtain the equation
\begin{equation}
\frac{du_x}{R(u_x+R)}=\frac{du^2}{2(3u_x+u^2 R)},
\label{wxw2Eq}
\end{equation}
which determines the functional connection between $u_x$ and $u^2$ in the
simple second sound wave with the velocity $\tilde V_1$ (\ref{V1}).

Eqs. (\ref{wxw2Eq}) can be integrated by changing variable $u^2$ to variable
$R$ (\ref{R}). After this substitution we get the differential
equation with separated variables
\begin{equation}
\frac{du_x}{1-u_x^2}=\frac{2R^2dR}{(R^2-3)(R^2-1)},
\label{wxREq}
\end{equation}
which can be immediately integrated. Its solution is
\begin{equation}
u_x(R)=\frac{\lambda-1}{\lambda+1},~~\lambda=C_{1}\left(
\frac{R-\sqrt{3}}{R+\sqrt{3}} \right)^{\sqrt{3}}\left(
\frac{R+1}{R-1} \right),
\label{wxR}
\end{equation}
where $C_1$ is an integration constant.

Solving Eq. (\ref{R}) with respect to $u^2$, we find
\begin{equation}
u^2(R)=1-2\frac{(1-u_x^2)}{R^2-1}.
\label{w2R}
\end{equation}
Relations (\ref{wxR}) and (\ref{w2R}) give, in parametric form, the functional
connection
between the $x$-component $u_x$, of the dimensionless relative
velocity ${\bf u}$, and its square
$u^2$, which corresponds to velocity
$\tilde V_1$, in the simple second sound wave
(\ref{V1}).

Substituting the first and second component of the eigenvector (\ref{r})
into the system of equations (\ref{ODEqs}), we obtain the other equation
\begin{equation}
d\Theta=\frac{(2u_{x}-(1-u^2)R)du_x}{(1-u^2)(u_x+R)R}.
\label{uwxEq}
\end{equation}

Expressing $u^2$ in terms of $R$, and $u_x$ by using Eq. (\ref{w2R})
in the factor
outside $du_x$ in Eq. (\ref{uwxEq}), we get  the following equation
instead of Eq. (\ref{uwxEq})
\begin{equation}
d\Theta-\frac{u_x du_x}{1-u_x^2}=-\frac{ du_x}{R(1-u_x^2)}.
\label{uwxEq2}
\end{equation}

Now, using Eq. (\ref{wxREq}) in  the right-hand side of Eq. (\ref{uwxEq2}),
we again get a differential equation with separated variables. By integrating
this equation, and then, taking into account the definition
(\ref{uDef}) of variable
$\Theta$, we obtain an expression for the temperature $T$ in the
simple second sound wave
\begin{equation}
T(R)=C_2 \sqrt{\frac{(1-u_x^2)}{(R^2-3)(R^2-1)}},
\label{TR}
\end{equation}
where $C_2$ is an integration constant.

Thus the relations (\ref{wxR}), (\ref{w2R}), and (\ref{TR}) express
in parametric
form the relationships between the $x$-component $u_x$ of the dimensionless
relative velocity ${\bf u}$, the square of the dimensionless relative
velocity $u^2$, and temperature $T$,
in the simple second sound wave, which propagates in "superfluid
vacuum" of $^4$He with the velocity $\tilde V_1$ (\ref{V1}).
The expression for the velocity $\tilde V_1$ (\ref{V1}) can be transformed
into a more simple form using (\ref{w2R})
\begin{equation}
\tilde V_{1}(R)=\frac{Ru_x-1}{R-u_x}.
\label{V1R}
\end{equation}
It follows from Eqs. (\ref{wxREq}), (\ref{w2R}), and (\ref{TR}), that $u_x$
and $u^2$ increase monotonically when $R$  increases, and $T$ decreases
monotonically when $R$ increases. Thus starting from any of the
variables  $u_x$,
  $u^2$, or $T$ at the initial moment of time we can determine the
corresponding initial
  function $R(x,t=0)$. The time evolution of the value of $R$ is
determined by the
running wave equation:
\begin{equation}
\frac{\partial R(x,t)}{c\partial t}+\tilde V_1(R)\frac{\partial
R(x,t)}{\partial x}=0,
\label{REq}
\end{equation}
where $\tilde V_1(R)$ is the velocity (\ref{V1R}).
The solution of Eq. (\ref{REq}) is well-known
\begin{equation}
x-c\tilde V_1(R)t=f(R),
\label{xR}
\end{equation}
where $f(R)$ is an arbitrary function.

The solution (\ref{xR}) shows that every value of $R$ runs with its
own velocity
$V_1(R)$ in the simple second sound wave.
The relations (\ref{xR}),(\ref{V1R})  along with the expressions
(\ref{wxR}), (\ref{w2R}), and (\ref{TR}),
  express in parametric
form the spatial dependence of the $x$-component $w_x=cu_{x}$ of the
relative velocity ${\bf w}$,
the square of the relative velocity $w^2=c^{2}u^2$, and temperature $T$
in the simple second sound wave at any moment of time.

It follows from Eq. (\ref{V1R}), that
$dV_1(R)/dR>0$. So it is clear that a simple second sound wave can, in
general, give rise to a
break in the continuous solution  . To determine the location and the
velocity of the break we should
return to the energy and momentum conservation laws, however we will
not be concerned
with that problem here.

The other simple second sound wave, which has the velocity $V_2$ (see
Eq. (\ref{V2})),  can be obtained from the solution (\ref{wxR}),
(\ref{w2R}), and (\ref{TR}),
which corresponds to the velocity $V_1$. For this purpose
  let us note that if $T(x,t)$, $w_x(x,t)$ and $w^2(x,t)$ is a
solution of the system
(\ref{uDef})-(\ref{MDef}), then  $T(-x,t)$, $-w_x(-x,t)$ and
$w^2(-x,t)$ is also a solution.
This transformation maps one simple second sound wave to the other.

\section{Expansion of a phonon pulse into the "phonon vacuum" of
superfluid helium}
To study the main features of the transverse expansion of phonon
pulses in superfluid
helium we will solve the following problem.
Let superfluid helium with ${\bf v}_s=0$ fill up
all space.
Let us consider the initial conditions
\begin{equation}
T(x,0)=\begin{cases} T_{0}, |x|<L_0 \\
0,  |x|>L_0 \\
\end{cases};~~w_x(x,0)=0;~~w^2(x,0)=w^2_0 ,
\label{InitCondAuto}
\end{equation}
for the system of Eqs. (\ref{uDef}), (\ref{MainEqs}) and (\ref{MDef}).

\begin{figure}[t]
\begin{center}
\includegraphics[height=2.5in]{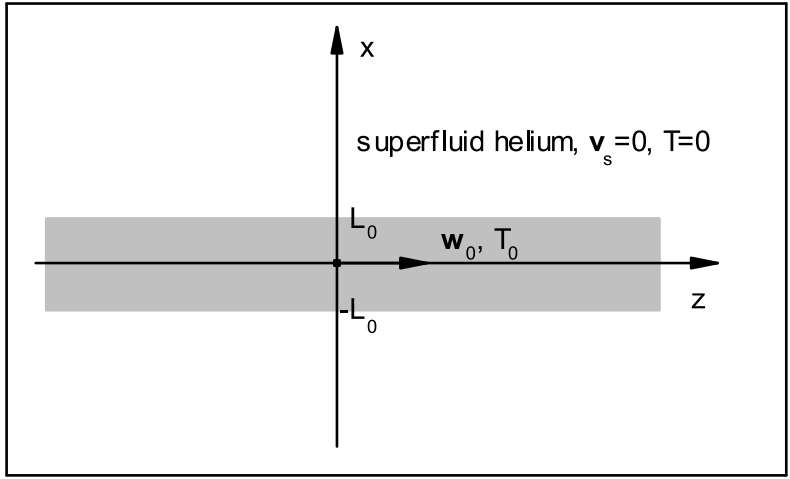}
\end{center}
\caption{A layer of phonons with temperature $T_0$ and the relative velocity
${\bf w}_0=w_0{\bf e}_z$  in superfluid helium with
  ${\bf v}_s=0$ and temperature $T=0$. The layer width is $2L_0$.
 The $x$-axis is directed perpendicularly to the plane of the layer.}
\label{fig2}
\end{figure}

At the initial time $t=0$, in the superfluid helium at $T=0$, there
is a layer of
phonons of width $2L_0$, temperature $T=T_{0}$, and relative
velocity ${\bf w}=(0,0,w_0)$ directed along $z$-axis (see Fig.\ref{fig2}).
The phonons in this initial layer is the normal fluid, and as the
phonon density is very low
the normal fluid density is very small
  so $\rho_n\ll\rho$. Therefore we can neglect the superfluid velocity and
pressure changes when we study the development of the phonon system.
Thus the set of Eqs. (\ref{MainEqs}) and (\ref{MDef}) apply to this
phonon system.
We are interested in finding the development in time of the
temperature $T$ and the relative velocity
${\bf w}$ of the phonon system.

The problem, formulated above, can be solved partly by using the
simple second
sound wave solution found in the previous
section.
First of all we note that the initial conditions
(\ref{InitCondAuto})
satisfy the symmetry transformations discussed
at the end of section 3. Therefore
at any time, the temperature $T$
and $w^2$ are even functions of  the $x$-coordinate,
and $w_x$ is an
odd function.  Below we will only discuss the solution for
domain
$x>0$.

\begin{figure}[t]
\begin{center}
\includegraphics[height=3in]{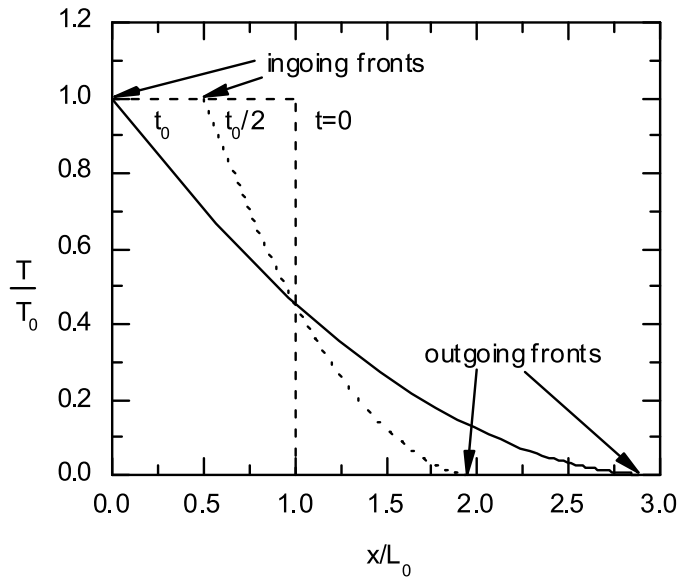}
\end{center}
\caption{The dependence of the
temperature $T$ on the x-coordinate for the initial value
$w_0=0.95c$
at time $t = 0$ (dashed line), $t = t_0/2$
(dotted line), and $t =
t_0$ (solid line).
The arrows point to the location
of the fronts of
the ingoing and outgoing waves at times $t = t_0/2$
and $t =
t_0$.}
\label{fig3a}
\end{figure}

The strong discontinuity, which exists at $x=L_0$ in the
initial conditions, disappears  immediately
after $t=0$, as it is
clear from physical considerations. The phonon gas expands into the
superfluid  helium with zero temperature, forming
the forward
front of the outgoing second sound   wave (see Fig.\ref{fig3a}). At the same
time,
 the initial perturbation, in the form of a weak discontinuity,
moves in the region $x<L_0$ forming the rear front of the ingoing
second sound   wave (see Fig.\ref{fig3a}). The weak discontinuity is at
the point where the temperature starts to change from its initial
value (\ref{InitCondAuto}), and it moves towards the coordinate
origin $x=0$ with the velocity  $c\tilde
V_1(u_x=0,u_{0}^2=w_0^2/c^{2})$ determined by Eq. (\ref{V1}). Thus
until the time $t_0$ when the intersection point reaches $x=0$,
there is no length scale in the problem, and the solution must be
self-similar. The value of $t_0$ is given by
\begin{equation}
t_{0}=\frac{L_0}{|V_1(u_x=0,u_{0}^2=w_0^2/c^{2})|}=\frac{L_0}{c}\sqrt{\frac{3-w_0^2}{1-w^2_0}},
\label{t0}
\end{equation}

A
self-similar solution is a
particular case
of a simple
wave.
Therefore from Eq. (\ref{xR}),
 taking into account that at the
initial time $t=0$ the wavefront is
at $x=L_0$, we get $f(R)=L_0$.
Thus the function $f(R)$  is reduced to constant for the self-similar
solution
and
\begin{equation}
x=L_{0}+V_1(R)t
\label{xRAuto}
\end{equation}
where
$V_1(R)=c\tilde V_{1}(R)$ is determined by the relation
(\ref{V1R}).

For any moment of time $0<t<t_0$, the wave occupies the
domain $x\in (x_{in}(t),x_{out}(t))$,
where value
$x_{in}(t)=L_{0}+V_1(R_0)t$ determines the position of the front of
the ingoing   wave, and
$x_{out}(t)=L_{0}+V_1(R_{out})t$ determines
the position of the front of the outgoing   wave,
which follows from
Eq. (\ref{xRAuto}) (see Fig.\ref{fig3a}).  We denote the yet-unknown
limits
of the variable $R$ as $R_0$ and $R_{out}$, i.e. $R\in
(R_0,R_{out})$.
Below we show that $R_{out}=+\infty,$ and that the
value
$R_0$ is determined by Eq. (\ref{R0})  coinciding with
the
value $R(u_x=0,u_{0}^2=w_0^2/c^{2})$ from Eq. (\ref{R}).

To
determine the integration constants, which are contained in Eqs.
(\ref{wxR}) and (\ref{TR}), we should join continuously
the simple
wave solution (\ref{wxR}), (\ref{w2R}) and (\ref{TR}) with the
to the
initial values (\ref{InitCondAuto}) at $R=R_0$ at the front of the
ingoing wave. Thus we
get the
solution
\begin{equation}
u_x(R)=\frac{\lambda-1}{\lambda+1},~\lambda=\left(
\frac{R_0-1}{R_0+1}\right)\left( \frac{R_0+\sqrt{3}}{R_0-\sqrt{3}}
\right)^{\sqrt{3}}\left( \frac{R+1}{R-1} \right)\left(
\frac{R-\sqrt{3}}{R+\sqrt{3}}
\right)^{\sqrt{3}},
\label{wxRAuto}
\end{equation}
\begin{equation}
T(R)=T_0
\sqrt{\frac{(R_0^2-3)(R_0^2-1)}{(R^2-3)(R^2-1)}(1-u_x^2)},
\label{TRAuto}
\end{equation}
and
the value
$R_0$
\begin{equation}
R_{0}=\sqrt{\frac{3-w_0^2/c^2}{1-w^2_0/c^{2}}}.
\label{R0}
\end{equation}

On
the front of the outgoing   wave, in accordance with the initial
conditions
(\ref{InitCondAuto}) the phonon  temperature is equal to
zero.
It follows from Eqs. (\ref{TRAuto}) and (\ref{wxRAuto}) that
this can only be satisfied
by $R_{out}=+\infty$.

Thus for any $t\in
(0,t_0)$, Eqs. (\ref{wxRAuto}), (\ref{TRAuto}),
(\ref{w2R}),
(\ref{xRAuto}), and (\ref{V1R})  determine the desired
self-similar solution of
the initial value problem  (\ref{MainEqs}),
(\ref{MDef}),
and (\ref{InitCondAuto}), and give the values of  $T$,
$w_x=cu_{x}$, and $w^2=c^{2}u^2$ at
any point
$x\in(x_{in}(t),x_{out}(t))$.

From Eq. (\ref{wxRAuto}) we find that
at the front of the outgoing  wave,
which borders with superfluid
helium at zero temperature, the $x$-component of
the dimensionless
relative velocity $u_x$ is equal
to
\begin{equation}
u_{xout}=u_x(R=+\infty)=\frac{\lambda_{out}-1}{\lambda_{out}+1},~\lambda_{out}=\left(
\frac{R_0-1}{R_0+1}\right)\left( \frac{R_0+\sqrt{3}}{R_0-\sqrt{3}}
\right)^{\sqrt{3}}.
\label{wxfAuto}
\end{equation}

From Eq.
(\ref{V1R}) it follows that this velocity $w_{xout}$ coincides
with
the velocity of the front of the outgoing
wave
\begin{equation}
V_1(R_{out}=+\infty)=w_{xout}=cu_{xout}.
\label{V1fAuto}
\end{equation}

Also
from Eq. (\ref{V1R}) in accordance with Eq. (\ref{t0}) for the
velocity of the front of the ingoing  wave, we
get
\begin{equation}
V_1(R_0)=-\frac{c}{R_0}=-c\sqrt{\frac{1-w^2_0}{3-w_0^2}},
\label{V1R0Auto}
\end{equation}
which
coincides with the velocity $c\tilde
V_1(u_x=0,u_{0}^2=w_0^2/c^{2})$
determined by Eq. (\ref{V1}).

It is
interesting that the relative velocity at the
front of the outgoing
wave, which follows
from Eq. (\ref{w2R}), is equal to the phonon
velocity i.e. $w^2(R_{out}=+\infty)=c^{2}$.
Whereas the temperature
of such phonons is equal to zero. Therefore such phonon
system
remains thermodynamically stable \cite{PhononStability}. The
phonon
energy density at the front of the outgoing   wave tends to
zero.

\begin{figure}[t]
\begin{center}
\includegraphics[height=3in]{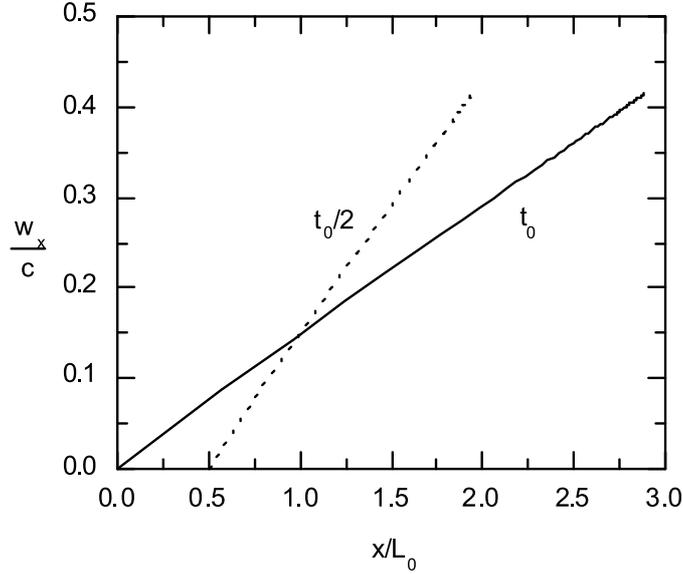}
\end{center}
\caption{The dependence of $x$-component
$w_x$ of the relative velocity
on the x-coordinate, for the initial
value
$w_0=0.95c$ for times $t = t_0/2$
(dotted line), and $t = t_0$
(solid line).}
\label{fig3b}
\end{figure}

\begin{figure}[t]
\begin{center}
\includegraphics[height=3in]{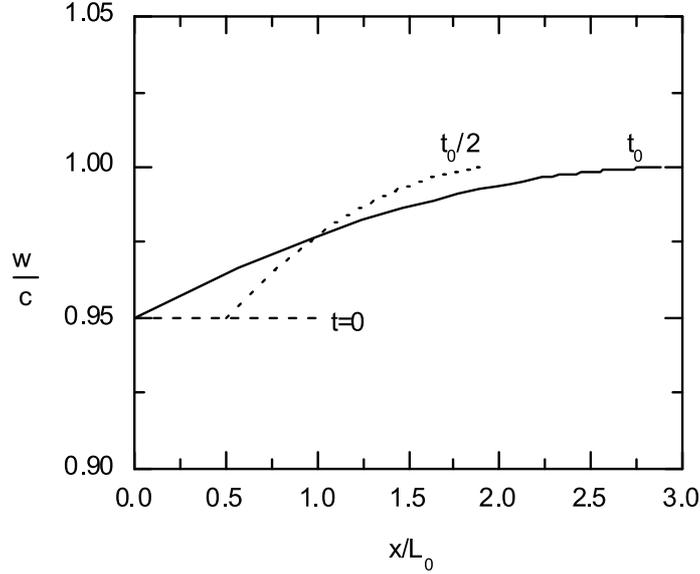}
\end{center}
\caption{The dependence of the
absolute value of the
relative velocity $w$ on the x-coordinate for
the initial value
$w_0=0.95c$ for times $t = 0$ (dashed line), $t =
t_0/2$
(dotted line), and $t = t_0$ (solid line).}
\label{fig3c}
\end{figure}

Figs.\ref{fig3a}, \ref{fig3b}, \ref{fig3c} show the spatial dependence of the temperature $T$
(Fig.\ref{fig3a} ), the $x$-component $w_x$ (Fig.\ref{fig3b})
of the relative
velocity ${\bf w}$, and the square of the relative velocity $w^2$
(Fig.\ref{fig3c})
for  times $t=0$ (dashed lines
on Figs.\ref{fig3a} and \ref{fig3c}),
$t=t_0/2$ (dotted lines in Figs.\ref{fig3a}, \ref{fig3b}, and \ref{fig3c}), and for $t=t_0$
(solid lines
in Figs.\ref{fig3a}, \ref{fig3b}, and \ref{fig3c}). These graphs are calculated
from Eqs. (\ref{wxRAuto}), (\ref{TRAuto}),
(\ref{w2R}),
(\ref{xRAuto}), and (\ref{V1R}) taking into account
(\ref{R0}) for the initial
value $w_0=0.95c$, which corresponds to a
strongly anisotropic phonon system.
We see that temperature (Fig.\ref{fig3a}) of the phonon pulse, expanding in "phonon vacuum",  decreases
monotonically,
but  the $x$-component $w_x$ (Fig.\ref{fig3b}) of the
relative velocity ${\bf w}$, and the square of the
relative velocity
$w^2$ (Fig.\ref{fig3c}) increase when $x$-coordinate increases.

Let us find
all the variables at the point $x=L_0$. If we substitute $x=L_0$
in
Eq. (\ref{xRAuto}), then we get the equation $V_1(R^\ast)=0$, where
we
denote as $R^\ast$ the value of $R$, which corresponds to $x=L_0$
at any time $t>0$.
Taking into account Eq. (\ref{V1R}), we obtain
the equivalent equation
$R^\ast u_x(R^\ast)=1$. Solving this equation
taking into account the expression
(\ref{wxRAuto}), we
get
\begin{equation}
R^\ast=\sqrt{3}\frac{1+\gamma}{1-\gamma},~\gamma=\left(
\frac{R_0-\sqrt{3}}{R_0+\sqrt{3}}\right)\left( \frac{R_0+1}{R_0-1}
\right)^{\frac{1}{\sqrt{3}}},
\label{Rast}
\end{equation}
where $R_0$
is determined by Eq. (\ref{R0}).

Substituting the value $R^\ast$,
determined by (\ref{Rast}), into Eqs. (\ref{wxRAuto}),
(\ref{TRAuto}), and (\ref{w2R}),
we obtain the temperature
$T(R^{\ast})$, the $x$-component $w_x(R^\ast)$  of the relative
velocity ${\bf w}$,
and the square of the relative velocity
$w^2(R^\ast)$ at $x=L_0$, i.e.  these are the values at the
initial
pulse boundary,  for any time while the solution is
self-similar. In particular we
get
\begin{equation}
u_x(R^\ast)=\frac{1}{R^\ast},~~u^2(R^{\ast})=1-\frac{2}{(R^{\ast
})^2}.
\label{wRast}
\end{equation}

If $R>R^\ast$, then $V_1(R)>0$,
therefore the wave  moves along
the positive direction of the $x$
axis, and if $R<R^\ast$ then vice versa.

\begin{figure}[t]
\begin{center}
\includegraphics[height=3in]{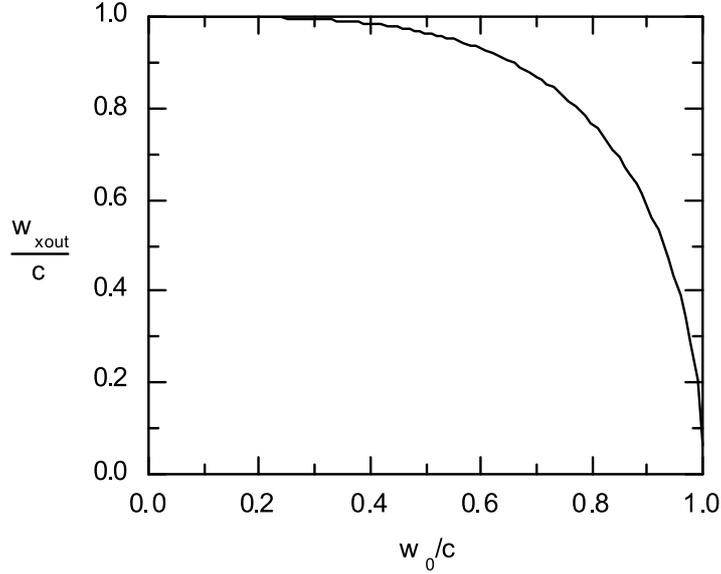}
\end{center}
\caption{The dependence of the expansion
velocity $w_{xout}$ of the phonon
pulse on the initial relative velocity $w_0$.}
\label{fig4}
\end{figure}

 Fig.\ref{fig4} shows the
dependence of the expansion velocity $w_{xout}$, of the phonon
pulse, on the initial relative velocity $w_0$, calculated from Eqs.
(\ref{wxfAuto}) and (\ref{R0}).
We see that for strongly anisotropic
phonon systems, when
$w_0 \sim c$, the expansion velocity $w_{xout}$
can be very small, compared
to the phonon velocity. When $w_0\ll c$,
then $w_{xout}\sim c$, and phonon
system expands nearly with the
phonon velocity.

The self-similar solution Eqs. (\ref{wxRAuto}),
(\ref{TRAuto}), (\ref{w2R}),
(\ref{xRAuto}), and (\ref{V1R}),
together with the unaffected region which remains constant  ,
determine the desired  solution of the initial value problem for Eqs.
(\ref{MainEqs}), (\ref{MDef}),
and (\ref{InitCondAuto}) everywhere only till moment of time $t_0$.
At this moment of time
the front of the ingoing  wave, which propagates in the region $x>0$, reaches
the coordinate origin $x=0$, and meets the front of the ingoing   wave  which
propagates in the region $x<0$ in a symmetric way to the $x>0$ wave.
At  $t>t_0$ these two waves
overlap at the centre region of the phonon pulse. Thus a reflected wave
arises in the region $0<x<x_{ref}(t)$, when $t>t_0$. In the region
$x_{ref}(t)<x<x_{out}(t)$ the solution remains  self-similar, and is
described by the
same formulae (\ref{wxRAuto}), (\ref{TRAuto}), (\ref{w2R}),
(\ref{xRAuto}), and (\ref{V1R}).  The point $x_{ref}(t)$ is determined from
the condition of a continuous join between the reflected wave and the
self-similar
solution.

Because we cannot find the analytical solution for the reflected wave, we
consider in the next section the approximate method for its description.

\section{Approximate description of  the reflected wave}
At the time $t=t_0$, when the front of the ingoing wave reaches
the coordinate origin $x=0$, the $x$-coordinate derivative of the
energy density
has a maximum at $x=0$, and the initial plateau on the energy density
curve has shrunk to zero.
After the time  $t=t_0$, when the reflected wave appears, it transfers
the energy from the coordinate origin, where the energy density is maximal,
  to the periphery, where the energy density tends to zero.  As a result,
in the region $0<x<x_{ref}(t)$   the energy density curve has an
approximate plateau,
  where the $x$-coordinate derivative of the energy density is small
compared to the one in the region
  where the reflected wave has not reached. This dependence  of the
energy density on the $x$-coordinate
(or on the angle between the normal to the heater and the detector)
has been experimentally
studied recently in Ref. \cite{Wyatt5}. At the present time we cannot
find the analytical solution for
the reflected wave because there are no general methods for solving
nonlinear equations
of the form (\ref{MainEqs}) and (\ref{MDef}).  However preliminary
numerical calculations confirm the picture presented here.

Starting from the condition that the reflected wave and the incident
wave join continuously, and using the energy conservation law, we can
find approximately the average
energy density and the average
width
of the reflected wave. For this purpose let us formulate the
energy conservation
law for one dimensional phonon propagation in
superfluid helium.

For a phonon
system with a linear energy-momentum
relation $\varepsilon=cp$, it is easy
to obtain the energy
density
\begin{equation}
E=\frac{\pi^2}{30}\frac{k_B^4T^4(1+w^2/(3c^{2}))}{\hbar^3c^3(1-w^2/c^{2})^3}.
\label{PhEDens}
\end{equation}
The
energy  conservation law can be written
as
\begin{equation}
\frac{\partial E}{\partial t}+ \frac{\partial
Q_E}{\partial x} =0,
\label{EEq}
\end{equation}
where $Q_E$ is the
phonon energy density flux in the
$x$-direction
\begin{equation}
Q_E=\frac{2\pi^2}{45}\frac{k_B^4T^4
w_x}{\hbar^3c^3(1-w^2/c^{2})^3}.
\label{PhEDensFlux}
\end{equation}

The
energy conservation law (\ref{EEq}) along with the expression for
the
energy density flux (\ref{PhEDensFlux}) can be derived directly
from the
system of equations (\ref{MainEqs}) and (\ref{MDef}), which
describe the propagation of the  phonon
system in superfluid
helium.

When the  phonon expansion is described by (\ref{MainEqs})
and (\ref{MDef}), together
with the initial conditions
(\ref{InitCondAuto}), the total energy, which
is localized at the
half-plane $x>0$, does not depend on time due to
the energy
conservation law (\ref{EEq}) and the symmetry of the initial
conditions (\ref{InitCondAuto}).
This energy is equal to $E_0 L_0$,
where
$E_0=E(T_0,w_0)=E(R_0)$ is the initial phonon energy density
(\ref{PhEDens}).

\begin{figure}[t]
\begin{center}
\includegraphics[height=3in]{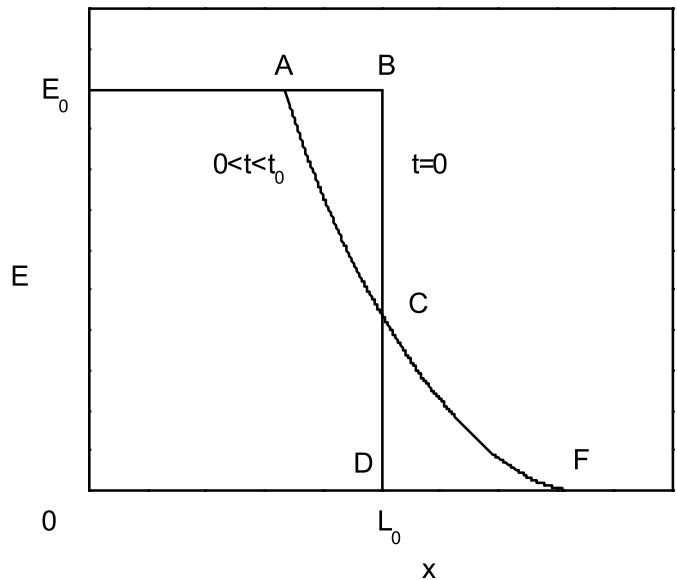}
\end{center}
\caption{The solid line $ACF$ presents the spatial dependence of the phonon
energy density $E$ at some time $0 < t < t_0$.}
\label{fig5}
\end{figure}

On Fig.\ref{fig5} the solid line  represents the spatial
dependence of the phonon energy density (\ref{PhEDens})
at some time
$0<t<t_0$, while the phonon pulse expansion is described
by the
self-similar solution (\ref{wxRAuto}), (\ref{TRAuto}),
(\ref{w2R}),
(\ref{xRAuto}), and (\ref{V1R}) (line $ACF$). The
energy conservation law leads to the equality
of areas $ABC$ and
$CDF$ in Fig.\ref{fig5}.

\begin{figure}[t]
\begin{center}
\includegraphics[height=3in]{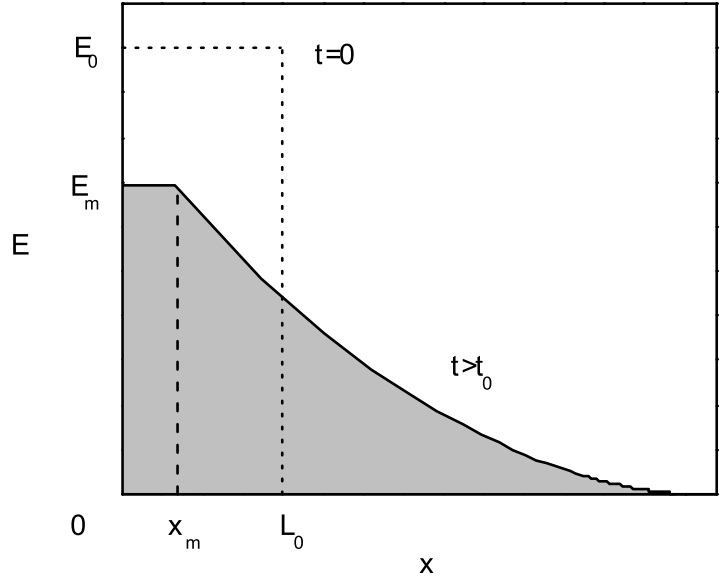}
\end{center}
\caption{The solid line presents the spatial dependence of the phonon
energy density $E$ at some time $t > t_0$. The values
$x_m$ and $E_m$ shows the approximate position and the average energy
density in the reflected wave respectively.}
\label{fig6}
\end{figure}

At the time $t>t_0$, in the central region
$0<x<x_{ref}(t)$, there is a reflected wave.
In the region
$x>x_{ref}(t)$ the solution is the self-similar wave.
To find
approximately the average energy density and the average width
of the
reflected wave
we  formally continue this self-similar solution into
the region $x<x_{ref}(t)$
until some point $x_m(t)$, which is
characterised by the energy density
$E_m(t)$. We show this in Fig.\ref{fig6},
where the self-similar wave is drawn
at some time $t>t_0$. This value
$E_m(t)$ along with $x_m(t)$ is determined from the
condition that
the  shaded area on Fig.\ref{fig6} is equal to the initial energy
$E_0 L_0$.
Thus we can write the equation
for $E_m$ and
$x_m$
\begin{equation}
E_m
x_m+\int_{R_{m}}^{+\infty}E(R)\frac{\partial x(R)}{\partial R}dR=E_0
L_0,
\label{ERm}
\end{equation}
where $x_m=x(R_m)$ (see Eq.
(\ref{xRAuto}))  and $E_m=E(R_m)$.

Substituting in Eq.
(\ref{ERm})
the expression (\ref{xRAuto}) for $x$-coordinate, we get
after  integration
by parts

\begin{equation}
t_m=\frac{(E(R_{m})-E_0)L_0}{\int_{R_{m}}^{+\infty}V_1(R)\frac{\partial
E(R)}{\partial R}dR}.
\label{tm}
\end{equation}
The expression
(\ref{tm}) determines approximately the time $t_m$, when the
reflected
wave reaches the value $R_m>R_0$. Note that the integral in
the denominator
of Eq. (\ref{tm}) tends to zero at $R_m=R_0$. This is
a mathematical expression
of the equality  of areas $ABC$ and $CDF$
in Fig.\ref{fig5} at any time
$t<t_0$.

It should be noted that if the energy
density gradient was zero in the
reflected wave region, then we would
have the equality $x_m(t)=x_{ref}(t)$
and Fig.\ref{fig6} would show the
exact  spatial dependence of the energy density, with the
characteristic
plateau in the central region. As
 it is clear
from a physical point of view and preliminary numerical calculations,
in the reflected wave region, the energy density gradient is much
smaller than the one in the ingoing wave,
but not equal to zero.
Therefore $x_m(t)$
does not coincide with $x_{ref}(t)$. In fact,
because  $dE/dx<0$,  $x_m(t)<x_{ref}(t)$.
But the difference  between
$x_m(t)$  and $x_{ref}(t)$ is small, if the energy density gradient
is small in the
reflected wave region compared to the one in the
ingoing wave. So the theory, developed above,
allows us to find the
average energy density and the average width
of the reflected wave.
This width  corresponds to the width of the mesa shape in the angular
distributions
observed in
\cite{Wyatt5}.

\begin{figure}[t]
\begin{center}
\includegraphics[height=3in]{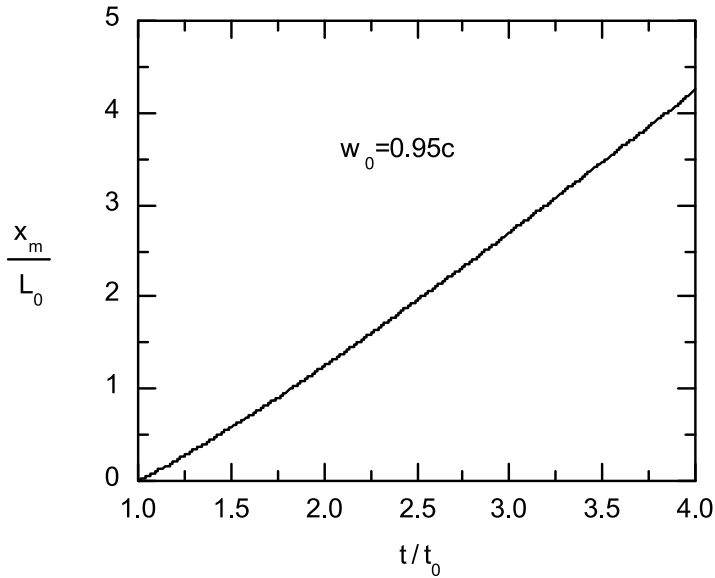}
\end{center}
\caption{The dependence of the approximate width $x_m$ of the reflected
wave on time $t$ for the initial value of $w_0=0.95c$.}
\label{fig7}
\end{figure}

Fig.\ref{fig7} shows the
dependence of the approximate width $x_m$ of the reflected wave on
time $t$,
calculated from Eqs. (\ref{tm}), (\ref{xRAuto}),
(\ref{PhEDens}), (\ref{wxRAuto}),
(\ref{TRAuto}), (\ref{w2R}) and
(\ref{V1R}). In
Fig.\ref{fig7} we see that this dependence is nearly linear.
This corresponds to the
constant velocity  of the front of the
outgoing self-similar wave.
In the experiments \cite{Wyatt5} it was
found that the mesa width increased with
the distance from the heater
which is the same as the dependence  of $x_m$  on time in 
Fig.\ref{fig7}.

\begin{figure}[t]
\begin{center}
\includegraphics[height=3in]{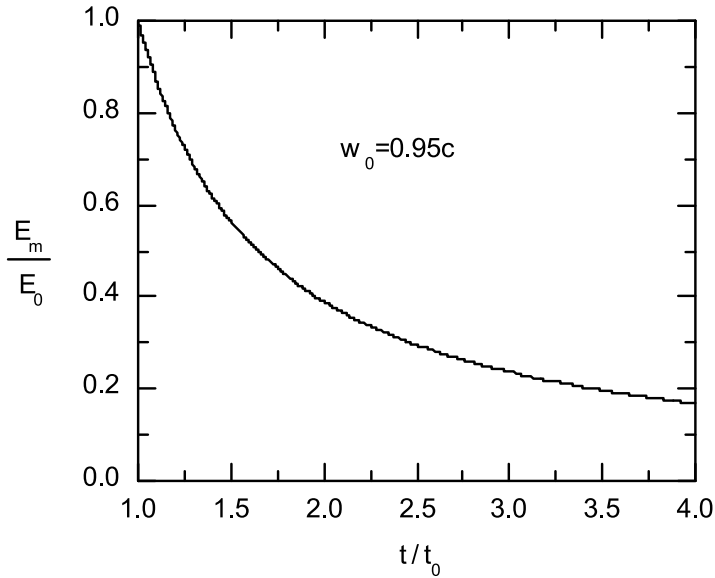}
\end{center}
\caption{The dependence of the approximate average energy density $E_m$, of the reflected
wave, on time $t$ for the initial value of $w_0=0.95c$.}
\label{fig8}
\end{figure}

The average energy density in the reflected wave $E_m$ as a
function of time $t$ is presented in Fig.\ref{fig8}.
It is calculated from
Eqs. (\ref{tm}), (\ref{PhEDens}), (\ref{wxRAuto}), (\ref{TRAuto}),
(\ref{w2R}), and (\ref{V1R}).  We see that the approximate average
energy density in the reflected
 wave $E_m$ decreases
monotonically
with time because of the expansion of the phonon pulse.
 The
dependence of $E_m$ on time
 (Fig.\ref{fig8}) is similar to that
observed in
experiments \cite{Wyatt5} where the mesa height decreased
with the
distance from the heater.

\section{Comparison with the experimental
data}
Initially, the phonon pulse  created by the heater (see Fig.\ref{fig1}),
 is comprised  of low energy phonons
(l-phonons), which
immediately start to create high energy phonons
(h-phonons)
\cite{HphCreation}.
The h-phonon creation results in
spatially smoothing the initial hot nonuniform  central region of
l-phonon pulse,
which is injected by heater into superfluid helium.
However the h-phonon creation for short
pulses occurs mainly near
the heater within a distance of several
millimeters
\cite{HphCreation}.
The h-phonon creation time
can  be
roughly estimated  as $10~\mu$s for the
typical experimental heater
powers. This time corresponds to the phonon travelling a distance
$2.4~$mm.
The initial conditions (\ref{InitCondAuto})
for the
l-phonon system correspond to the end of the h-phonon creation stage
of the evolution of the phonon system.
So the above theory is for the
expansion of a cool l-phonon system.

After the h-phonon creation rate has become negligible, the
temperature is $T_0$ and the relative velocity is $w_0$. The
values of $T_0$ and $w_0$ depend on the power applied to the
heater. Their values for l-phonon pulses, with a linear
momentum-energy relation, are calculated using the results of
\cite{HphCreation} for experimentally applied powers.
 The results of
these calculations are presented in Table 1
for the heater powers
used
in the experiments \cite{Wyatt5}. The second and third columns
show the corresponding
values of temperature $T_0$ and the relative
velocity $w_0$ for these powers.
Using the theory presented in this
paper, we can find
at any time the spatial dependencies of all the
parameters  of the l-phonon system for different powers,
which
correspond to the values of $T_0$ and  $w_0$ in Table 1.

\begin{table}
\begin{center}
\begin{tabular}{|c|c|c|}
\hline
Power
&
\multicolumn{2}{c|}{Initial values}\\
 \hline
$P$, mW&    $T_0$, K&
$w_0/c$ \\
\hline

3.125&  0.033& 0.955 \\

\hline
6.25& 0.041& 0.947 \\

\hline
12.5&  0.050& 0.938 \\

\hline
25& 0.065& 0.921 \\
\hline

\end{tabular}
\end{center}
\caption{The values of temperature $T_{0}$ and the relative velocity $w_{0}$
(divided by first sound velocity $c=238$ m/s) for the l-phonon
system  with different
values of the heater power P.}
\label{Table1}
\end{table}

We should
note that the   phonon pulses created in experiments
\cite{Wyatt5}
are not long enough to neglect the  $z$-coordinate
dependence in the phonon evolution
system (\ref{SEq}), (\ref{wEq}).
Moreover the experimental conditions are
much more close to
cylindrical symmetry, then to a plane one.  Nevertheless
we think
that the qualitative dependencies, obtained here for  a plane
phonon
layer, can be compared to  the real much more complicated
experimental conditions.

The theory presented here shows that the
initial phonon pulse  starts  to expand into the "phonon vacuum". The simple second sound
wave appears with the outgoing wave front
moving into the "phonon vacuum", and the front of the ingoing wave moving to the centre at $x=0$ (see Fig.\ref{fig3a}). During this time, the
width of initial plateau  decreases with time.
So at the time  $t_0$
(see Eq. (\ref{t0})) there is no mesa. The phonon pulse propagates a
distance of
$l_0 = c t_0$ during this time. For
the initial pulse
width $2L_0 = 1~$mm
and for powers  3.125 mW, 6.25 mW, 12.5  mW,  and
25 mW from Ref. \cite{Wyatt5}, we obtain the values of
$l_0$ of 2.4
mm, 2.3 mm, 2.1 mm, and 1.9 mm respectively. The total distance
between the heater and
the phonon pulse $l_h$ is $ l_h=l_0+l_c$,
where $l_c=2.4~$mm is the length after which the h-phonon
creation
has effectively stopped. Note that this schema of taking into account of h-phonon creation is rather rough one, especially for high powers.

   In experiments \cite{Wyatt5} the
minimal distance between heater and detector was 8.2 mm.
Thus  the decrease in the width of the initial plateau is not observed in
\cite{Wyatt5}.  It would need measurements of the
phonon energy
angular distributions at a distance about 3-4 mm from the heater to
compare with this
prediction of the  theory that there is no mesa at
$\approx 4$ mm.

At time $t=t_0$ the reflected wave appears.
From a
physical point of view it is clear that in the reflected wave,
which
occurs in the central region of the phonon pulse, the energy
density gradient
is smaller than in the ingoing wave. Thus the
existence of the reflected wave should
lead to  a mesa shape in the
phonon angular distributions at distances larger than $\approx 4$ mm.

In Ref.\cite{Wyatt5} the  detailed measurements of
l-phonon
angular distributions in superfluid helium showed a distinct mesa
shape.
The dependence of the mesa height and
width,  on heater
dimensions, distance to the detector
and heater power, were measured
\cite{Wyatt5}. Using the theory presented in this paper
we now can give qualitative explanations some of these experimental data.

\begin{figure}[t]
\begin{center}
\includegraphics[height=3in]{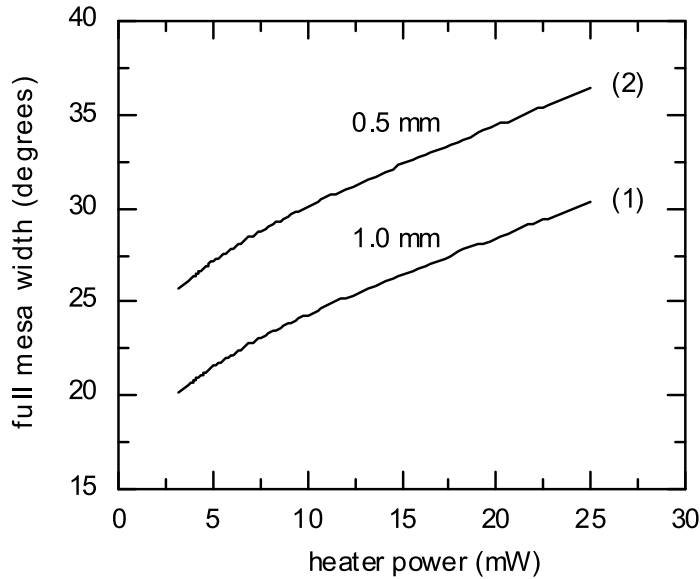}
\end{center}
\caption{The dependencies of the full mesa width $2x_m$ in degrees, on heater
power for the
distance  12.3 mm from the heater, for the heater widths
$2L_0 = 1~$mm
(line 1) and for $2L_0 = 0.5~$mm (line 2).}
\label{fig9}
\end{figure}

In Fig.\ref{fig9}
we show the dependencies of mesa width $x_m$  on power for  the
distance 12.3 mm
from the heater for the heater widths  $2L_0=1~$mm
(line 1) and for $2L_0=0.5~$mm
(line 2), calculated from the data in
Table 1 and Eqs. (\ref{tm}),
(\ref{xRAuto}), (\ref{wxRAuto}),
(\ref{V1R}), (\ref{R0}) and(\ref{w2R}). Here and below we
take  the
distances and the heater widths which were used in the  experiments
\cite{Wyatt5}.
  We see the monotonic growth of the mesa width $x_m$,
when power increases as it does in
the experimental data from
\cite{Wyatt5}, although the measured mesa widths are considerably smaller than those theoretically predicted. The power dependence can be explained in the following way.
When power increases, the
relative velocities $w_0$ decrease (Table 1). This results in increasing the absolute value of velocity
$V_1$. Thus the time $t_0$ (see Eq. (\ref{t0})) decreases, and the
mesa forms earlier. It should be noted that in the experiments \cite{Wyatt5} for some cases the measured mesa widths start to decrease at high power 25 mW (see Fig. 9 of \cite{Wyatt5}). Our theory, which describes the expansion of a cool l-phonon layer, cannot explain this effect.

\begin{figure}[t]
\begin{center}
\includegraphics[height=3in]{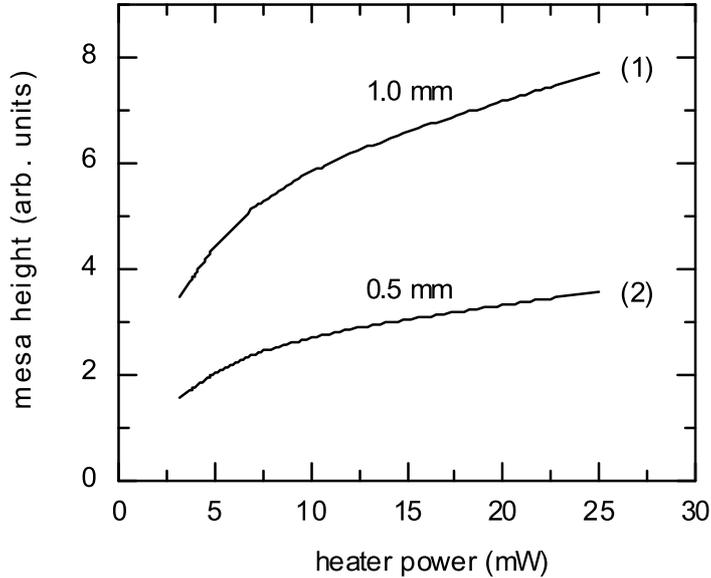}
\end{center}
\caption{The dependencies of the mesa height $E_m$ on power for the
distance 12.3 mm from the heater for the heater widths
$2L_0 = 1~$mm
(line 1) and for $2L_0 = 0.5~$mm (line 2).}
\label{fig10}
\end{figure}

In Fig.\ref{fig10} we present the dependencies
of the
mesa height $E_m$ on power for the distance 12.3 mm
from the heater
for the heater widths  $2L_0=1~$mm (line 1) and for
$2L_0=0.5~$mm
(line 2), calculated from the data in Table 1 and Eqs.
(\ref{tm}), (\ref{wxRAuto}),
(\ref{TRAuto}), (\ref{V1R}), (\ref{R0})
and(\ref{w2R}).
The mesa height $E_m$
increases, when the power
increases, as it does in the measured data \cite{Wyatt5}.

While the
mesa width $x_m$ do not depend on the initial temperature $T_0$, which follows from Eqs. (\ref{tm}) and (\ref{xRAuto}), the mesa
height $E_m$ depends on $T_0^4$.
It turns out that the increase
of
$T_0$ with power prevails over the decrease the relative velocity
$w_0$,
and so an increase in power leads to a quicker
expansion of
the phonon pulse and hence to a wider mesa.

The increase of $x_m$
with time,  shown in Fig.\ref{fig7}
 corresponds to  an increase of the mesa
width with distance. This is observed in experiments
\cite{Wyatt5}.
The decrease of  $E_m$ with time, shown in Fig.\ref{fig8},
corresponds to a decrease in the mesa height with distance.
This too
is observed in experiments \cite{Wyatt5}.

\begin{figure}[t]
\begin{center}
\includegraphics[height=3in]{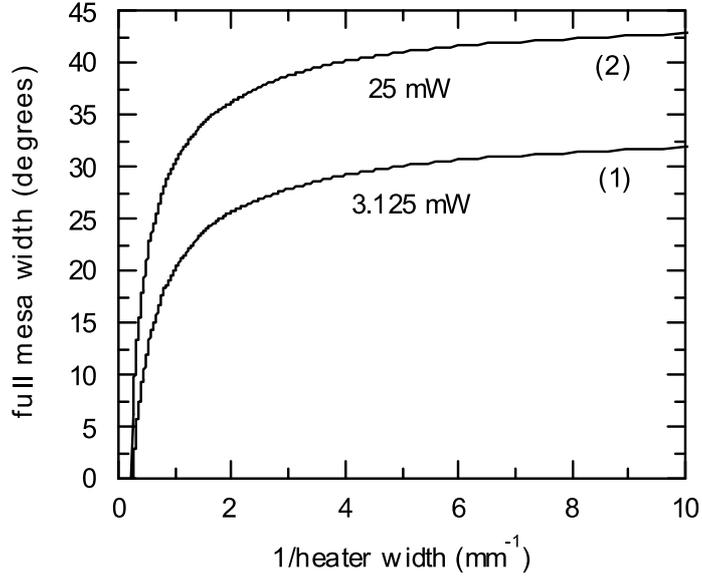}
\end{center}
\caption{The angular width of the full mesa  $2x_m$ versus the inverse of the heater
width for the distance 12.3 mm from the heater, for heater powers
$3.125~$mW (line 1) and $25~$mW (line 2).}
\label{fig11}
\end{figure}

The theory developed in
this paper allows us     to qualitatively understand why the  mesa
width increases
with decreasing the heater width \cite{Wyatt5}. A
smaller heater width ($2L_0$) results in an earlier
formation of the
mesa, and this leads directly to an increase in    the mesa width.
In
Fig.\ref{fig11} we  show the dependencies of mesa width $x_m$  on reciprocal
heater
width $1/(2L_0)$ for  the distance 12.3 mm
from the heater for
powers $3.125~$mW (line 1) and $25~$mW
(line 2), calculated from the
data in Table 1 and Eqs. (\ref{tm}), (\ref{xRAuto}), (\ref{wxRAuto}),
(\ref{V1R}), (\ref{R0}) and (\ref{w2R}). But we should note that our theory does not reproduce the linear dependence of mesa width on reciprocal heater width, which was observed in the experiments \cite{Wyatt5} (see Fig. 10 in \cite{Wyatt5}).

\section{Conclusions}
In this paper, starting from the hydrodynamic equations of superfluid helium,
we have obtained  the system of equations (\ref{SEq}) and
(\ref{wEq}), which describe the  evolution
of a cool phonon systems, created by thermal pulses  in superfluid
helium for the case $\rho_n\ll\rho$ (see Fig.\ref{fig1}).
These equations are simplified to the case of one spatial dimension
(see Eqs.  (\ref{uDef})-(\ref{MDef}), and Fig.\ref{fig2}).  The family of
exact solutions
were found in an explicit analytical form. They are the
simple waves of second sound for phonon systems. The relations
between temperature $T$ (\ref{TR}),
the $x$-component $w_x$ (\ref{wxR}) of the relative velocity ${\bf w}$, and
the square of the relative velocity $w^2$ (\ref{w2R}) are studied for
the simple second sound
waves in phonon systems. These solutions are used
to describe the first stage of expansion of a phonon  layer in superfluid
helium, when only simple second sound waves exist. Figs.\ref{fig3a}, \ref{fig3b},
and \ref{fig3c} show the
spatial dependence of temperature $T$ (Fig.\ref{fig3a}),
the $x$-component $w_x$ (Fig.\ref{fig3b}) of the relative velocity ${\bf w}$,
and the square of the
relative velocity $w^2$ (Fig.\ref{fig3c}) for time $t=0$ (dashed lines
on Figs.\ref{fig3a} and \ref{fig3c}),  $t=t_0/2$ (dotted lines in Figs.\ref{fig3a},
 \ref{fig3b}, and
\ref{fig3c}), and for $t=t_0$ (solid lines
in Figs.\ref{fig3a}, \ref{fig3b}, and \ref{fig3c}).

We have found the velocity of expansion $w_{xout}$ of a phonon pulse,
propagating
initially in $z$-direction, into the "phonon vacuum", i.e. into
superfluid helium at zero temperature.
The dependence of the expansion velocity $w_{xout}$ of the phonon
  pulse, on the initial relative velocity $w_0$, calculated from Eqs.
(\ref{wxfAuto}) and (\ref{R0})
are presented in Fig.\ref{fig4}. We see that for strongly anisotropic phonon
systems, when
$w_0 \sim c$, the expansion velocity $w_{xout}$ can be small compared
to the phonon velocity. When $w_0\ll c$, then $w_{xout}\sim c$, and phonon
system expands nearly with the phonon velocity.

In the second stage, after the incident wave reaches the centre of
the phonon layer, a reflected wave appears.
We found the time $t_0$ Eq. (\ref{t0}), when this starts.
The reflected wave  transfers
energy from the coordinate origin, where the energy density is maximal,
  to the periphery, where the energy density tends to zero. It smooths
the
dependence of the energy density on  the $x$-coordinate, in the
reflected wave
region.
The smallness of  the $x$-coordinate derivative of the energy density
in the reflected wave
region, compared with the one in the ingoing wave, results in a mesa
shape form,
which was observed in \cite{Wyatt5}.

We developed an approximate
theory for the average energy density and the average width
of the reflected wave (see Fig.\ref{fig6}).  The calculated dependencies of the
mesa height  and mesa width
on time (Figs.\ref{fig7}, \ref{fig8}), on the heater power 
(Figs.\ref{fig9}, \ref{fig10}), and on the heater
width (Fig.\ref{fig11}) show partly the same qualitative dependencies as the
experimental data in \cite{Wyatt5}, although our theory fails to explain all effects observed. But we think that the main cause of the mesa shape appearance in the experiments \cite{Wyatt5} are the same as in our simple model theory.

\section*{Acknowledgements}
We thank very much A. F.G. Wyatt for drawing our attention to this problem
and for many useful discussions,
and we express our gratitude to EPSRC of the UK (grant EP/F 019157/1)
for support of this work.

\end{document}